\documentclass[12pt, a4paper]{article}
\usepackage{geometry, amsmath,amssymb,amsthm,amsxtra,overpic,bbm,bm,epsfig,sub
figure}
\usepackage{mathrsfs}
\usepackage{graphicx}
\usepackage{color}
\usepackage{comment}
\usepackage{epstopdf}
\usepackage{float}

\usepackage{slashed,stmaryrd}

\usepackage{multirow}

\def\thefootnote{\fnsymbol{footnote}}

\begin{document}

\vspace{0.2cm}

\begin{center}
{\large\bf Distinguishing between the twin $b$-flavored unitarity triangles \\
on a circular arc}
\end{center}

\vspace{0.2cm}

\begin{center}
{\bf Zhi-zhong Xing$^{1,2}$}
\footnote{E-mail: xingzz@ihep.ac.cn}
and {\bf Di Zhang$^{1}$}
\footnote{E-mail: zhangdi@ihep.ac.cn (corresponding author)}
\\
{\small $^{1}$Institute of High Energy Physics and School of Physical Sciences, \\
University of Chinese Academy of Sciences, Beijing 100049, China \\
$^{2}$Center of High Energy Physics, Peking University, Beijing 100871, China}
\end{center}

\vspace{2cm}
\begin{abstract}
With the help of the generalized Wolfenstein parametrization of quark flavor mixing
and CP violation, we calculate fine differences between the twin $b$-flavored
unitarity triangles defined by $V^{*}_{ub} V^{}_{ud} +
V^{*}_{cb} V^{}_{cd} + V^{*}_{tb} V^{}_{td} = 0$ and $V^{*}_{ud} V^{}_{td}
+ V^{*}_{us} V^{}_{ts} + V^{*}_{ub} V^{}_{tb} = 0$ in the complex plane. We find that
apexes of the rescaled versions of these two triangles, described respectively
by $\overline{\rho} + {\rm i} \overline{\eta} =
-\left(V^{*}_{ub} V^{}_{ud}\right)/\left(V^{*}_{cb} V^{}_{cd}\right)$ and
$\widetilde{\rho} + {\rm i} \widetilde{\eta} =
-\left(V^{*}_{ub} V^{}_{tb}\right)/\left(V^{*}_{us} V^{}_{ts}\right)$,
are located on a circular arc whose center and radius are given by
$O = \left(0.5, 0.5 \cot\alpha\right)$ and $R = 0.5 \csc\alpha$ with $\alpha$
being their common inner angle. The small difference between $(\overline{\rho},
\overline{\eta})$ and $(\widetilde{\rho}, \widetilde{\eta})$ is characterized by
$\widetilde{\rho} - \overline{\rho} \sim \widetilde{\eta} - \overline{\eta}
\sim {\cal O}(\lambda^2)$ with $\lambda \simeq 0.22$ being the Wolfenstein expansion
parameter, and these two apexes are insensitive to the two-loop renormalization-group
running effects up to the accuracy of ${\cal O}(\lambda^4)$. Some comments are
also made on similar features of three pairs of the rescaled unitarity triangles
of lepton flavor mixing and CP violation.
\end{abstract}

\def\thefootnote{\arabic{footnote}}
\setcounter{footnote}{0}

\newpage

\section{Introduction}

The running Belle II detector at the KEK super-$B$ factory \cite{Kou:2018nap}
and an upgrade of the LHCb detector at CERN \cite{Bona:2007qt,Bediaga:2018lhg}
are pushing $B$-meson physics and CP violation into a new era, in
which further precision measurements will be done to test the standard model (SM)
predictions for $b$-flavored hadrons to an unprecedented degree of accuracy
and probe possible new physics via their fine quantum corrections to the SM results.
An intriguing question is whether these two experiments can finally distinguish
between the following {\it twin} $b$-flavored unitarity triangles of the
Cabibbo-Kobayashi-Maskawa (CKM) quark flavor mixing matrix $V$
\cite{Cabibbo:1963yz,Kobayashi:1973fv} in the complex plane:
\begin{eqnarray}
\triangle^{}_s ~: && V^{*}_{ub} V^{}_{ud} + V^{*}_{cb} V^{}_{cd} +
V^{*}_{tb} V^{}_{td} = 0 \; ,
\nonumber \\
\triangle^{}_c ~: && V^{*}_{ud} V^{}_{td} + V^{*}_{us} V^{}_{ts} +
V^{*}_{ub} V^{}_{tb} = 0 \; ,
\end{eqnarray}
where each triangle is named after the flavor index that does not appear
in its three sides \cite{Fritzsch:1999ee,Xing:2019vks}. The reason is simply
that these two triangles are almost congruent with each other, and hence their
differences in sides and inner angles are experimentally indistinguishable
at present. Notice that the triangle $\triangle^{}_s$ has played a crucial role
in verifying the success of the Kobayashi-Maskawa mechanism of CP violation,
and it has been extensively studied in various decays of $B^{}_d$ and $B^{}_u$
mesons in the past twenty years.
In comparison, the triangle $\triangle^{}_c$ seems to have a much closer
relationship with the $B^{}_s$-meson decays, so its sides and inner angles
are expected to be carefully measured in the next twenty years at the LHCb
and Belle II experiments. Needless to say, an
experimental confirmation of high similarity between $\triangle^{}_c$ and
$\triangle^{}_s$ will provide us with stronger evidence for correctness
of the SM in the quark flavor mixing sector, while a discovery of some
unexpected and unsuppressed differences between these twin triangles will serve
as a remarkable signal of new physics beyond the SM. Such a point was already
emphasized by Bigi and Sanda two decades ago \cite{Bigi:1999hr}, and {\it now it is
time for us to confront it with the upcoming measurements at the
super-$B$ factory and the high-luminosity Large Hadron Collider (LHC)}.

The main purpose of this work is two-fold. On the one hand, we are going to
carry out a careful study of fine differences
between the {\it rescaled} versions of $\triangle^{}_s$ and $\triangle^{}_c$,
\begin{eqnarray}
\triangle^{\prime}_s ~: && 1 + \frac{V^{*}_{ub} V^{}_{ud}}{V^{*}_{cb} V^{}_{cd}}
+ \frac{V^{*}_{tb} V^{}_{td}}{V^{*}_{cb} V^{}_{cd}} = 0 \; ,
\nonumber \\
\triangle^{\prime}_c ~: && 1 + \frac{V^{*}_{ud} V^{}_{td}}{V^{*}_{us} V^{}_{ts}}
+ \frac{V^{*}_{ub} V^{}_{tb}}{V^{*}_{us} V^{}_{ts}} = 0 \; ,
\end{eqnarray}
with the help of the Wolfenstein parametrization of $V$
\cite{Wolfenstein:1983yz}. On the other hand, we shall examine the stability of
$\triangle^{\prime}_s$ and $\triangle^{\prime}_c$ against quantum
corrections by using the two-loop renormalization-group equations (RGEs) of
the CKM matrix $V$ \cite{Machacek:1983fi,Barger:1992ac,Barger:1992pk,Luo:2002ey}.
Analogous to the definition
\begin{eqnarray}
\overline{\rho} + {\rm i} \overline{\eta} =
-\frac{V^{*}_{ub} V^{}_{ud}}{V^{*}_{cb} V^{}_{cd}} \;
\end{eqnarray}
for the apex of $\triangle^{\prime}_s$ \cite{Buras:1994ec,Charles:2004jd,
Tanabashi:2018oca}, the apex of $\triangle^{\prime}_c$ can be defined as
\begin{eqnarray}
\widetilde{\rho} + {\rm i} \widetilde{\eta} =
-\frac{V^{*}_{ub} V^{}_{tb}}{V^{*}_{us} V^{}_{ts}} \;
\end{eqnarray}
in the complex plane. The analytical expressions of $(\overline{\rho}, \overline{\eta})$
and $(\widetilde{\rho}, \widetilde{\eta})$ will be derived up to the accuracy of
${\cal O}(\lambda^6)$, where $\lambda \simeq 0.22$ is the Wolfenstein expansion
parameter, so as to see the tiny difference between these two rescaled unitarity
triangles. We find that the apexes $(\overline{\rho}, \overline{\eta})$
and $(\widetilde{\rho}, \widetilde{\eta})$ are actually located on a circular arc in the
complex plane. Both $\widetilde{\rho} - \overline{\rho}$ and
$\widetilde{\eta} - \overline{\eta}$ are of ${\cal O}(\lambda^2)$, and thus
$\triangle^\prime_c$ and $\triangle^\prime_s$ should be experimentally distinguishable
at the $3\sigma$ level if their apexes $(\overline{\rho} , \overline{\eta})$ and
$(\widetilde{\rho} , \widetilde{\eta})$ can be measured to the $\lesssim 0.4\%$
precision. The similar experimental sensitivity will allow us to probe
small differences between the inner angles of $\triangle^\prime_c$ and
$\triangle^\prime_s$. A possible way to separately establish $\triangle^\prime_s$ and $\triangle^\prime_c$ is to use the experimental data from $B^{\pm}_u$ and
$B^0_d$-$\bar{B}^0_d$ systems and those from $B^{\pm}_u$ and $B^0_s$-$\bar{B}^0_s$
systems, respectively. We also find that $(\overline{\rho}, \overline{\eta})$
and $(\widetilde{\rho}, \widetilde{\eta})$ are insensitive to the two-loop RGE
running effects up to the accuracy of ${\cal O}(\lambda^4)$, implying that
the shapes of $\triangle^{}_c$ and $\triangle^{}_s$ keep invariant up to the
same accuracy when the energy scale evolves from the
electroweak scale $\Lambda^{}_{\rm EW} \sim 10^2~{\rm GeV}$ to the scale of a
grand unification theory (GUT) --- $\Lambda^{}_{\rm GUT} \sim 10^{16}~{\rm GeV}$
(or vice versa). Therefore, the experimental results of all the inner angles of
$\triangle^{}_c$ and $\triangle^{}_s$ obtained at low energies can directly be
confronted with some theoretical predictions at a superhigh energy scale.
We finally make some brief comments on similar features of three pairs of the
rescaled unitarity triangles of lepton flavor mixing and CP violation in the
complex plane, which are expected to be useful for the study of neutrino oscillations
in the near future.

\section{The apexes of $\triangle^\prime_s$ and $\triangle^\prime_c$
on a circular arc}

Let us start from a popular extension of the original Wolfenstein parametrization of
the CKM matrix $V$ proposed by Buras {\it et al} in 1994 \cite{Buras:1994ec,Charles:2004jd,
Tanabashi:2018oca}
\footnote{The point of this extension is to start from the exact ``standard" (Euler-like)
parametrization of $V$ \cite{Tanabashi:2018oca} to expand each element of $V$ by
defining $\sin\vartheta^{}_{12} \equiv \lambda$, $\sin\vartheta^{}_{23} \equiv A \lambda^2$
and $\sin\vartheta^{}_{13} e^{{\rm i}\delta} \equiv A\lambda^3 \left(\rho + {\rm i}\eta\right)$,
where $\vartheta^{}_{ij}$ (for $ij = 12, 13, 23$) are the flavor mixing angles of
six quarks and $\delta$ is the irreducible CP-violating phase of the CKM matrix $V$ in
this parametrization. Such a treatment, which is slightly different from the convention of
$V^{}_{us} \equiv \lambda$, $V^{}_{cb} \equiv A \lambda^2$ and $V^{}_{ub} \equiv A\lambda^3 \left(\rho -{\rm i} \eta\right)$ taken originally by Wolfenstein \cite{Wolfenstein:1983yz}
and later by Kobayashi \cite{Kobayashi:1994ps}, has proved to be more convenient
for the study of heavy flavor physics.}.
Since ${\cal O}(\lambda^6)$ is equivalent to
${\cal O}(10^{-4})$, it is good enough for us to expand each element of $V$ in
powers of $\lambda$ up to this degree of accuracy. To be explicit, one actually finds
\begin{eqnarray}
&& V^{}_{ud} = 1 - \frac{1}{2}\lambda - \frac{1}{8}\lambda^4 + {\cal O}(\lambda^6) \; ,
\nonumber \\
&& V^{}_{us} = \lambda + {\cal O}(\lambda^7) \; ,
\nonumber \\
&& V^{}_{ub} = A\lambda^3 \left(\rho - {\rm i}\eta\right) \; ;
\nonumber \\
&& V^{}_{cd} = -\lambda + \frac{1}{2} A^2 \lambda^5 \left[1 - 2\left(\rho + {\rm i}
\eta\right)\right] + {\cal O}(\lambda^7) \; ,
\nonumber \\
&& V^{}_{cs} = 1 - \frac{1}{2} \lambda^2 - \frac{1}{8} \lambda^4 \left(1 + 4 A^2\right)
+ {\cal O}(\lambda^6) \; ,
\nonumber \\
&& V^{}_{cb} = A \lambda^2 + {\cal O}(\lambda^8) \; ;
\nonumber \\
&& V^{}_{td} = A\lambda^3 \left(1 -\rho - {\rm i}\eta\right) + \frac{1}{2}
A\lambda^5 \left(\rho + {\rm i}\eta\right) + {\cal O}(\lambda^7) \; ,
\nonumber \\
&& V^{}_{ts} = -A\lambda^2 + \frac{1}{2} A\lambda^4 \left[1 - 2\left(\rho +
{\rm i}\eta\right)\right] + {\cal O}(\lambda^6) \; ,
\nonumber \\
&& V^{}_{tb} = 1 - \frac{1}{2} A^2\lambda^4 + {\cal O}(\lambda^6) \; .
\end{eqnarray}
Note that here $V^{}_{ub} = A \lambda^3 \left(\rho - {\rm i}\eta\right)$ is exact
by definition. Substituting the above expressions into Eq.~(3) and Eq.~(4),
we directly arrive at
\begin{eqnarray}
&& \overline{\rho} = \rho \left\{ 1 - \frac{1}{2} \lambda^2 -
\left[\frac{1}{8} - \left(\frac{1}{2} -\rho + \frac{\eta^2}{\rho}\right)
A^2\right] \lambda^4 \right\} + \mathcal{O}(\lambda^6) \; ,
\nonumber \\
&& \overline{\eta} = \eta \left\{ 1 - \frac{1}{2} \lambda^2 -
\left[ \frac{1}{8} - \left(\frac{1}{2} -2\rho\right) A^2 \right]
\lambda^4 \right\} + \mathcal{O}(\lambda^6) \; ;
\end{eqnarray}
and
\begin{eqnarray}
&& \widetilde{\rho} = \rho \left\{ 1 + \left(\frac{1}{2} -\rho +
\frac{\eta^2}{\rho} \right) \lambda^2 + \left[ \frac{3}{8} - \frac{1}{2} A^2
-\rho \left(1 - \rho\right) - 3\eta^2 + \frac{\eta^2}{\rho} \right] \lambda^4
\right\} + \mathcal{O}(\lambda^6) \; ,
\nonumber \\
&& \widetilde{\eta} = \eta \left\{ 1 + \left(\frac{1}{2} - 2\rho \right)
\lambda^2 + \left[ \frac{3}{8} - \frac{1}{2} A^2 - \eta^2 - \left(2 - 3\rho\right)
\rho \right] \lambda^4 \right\} + \mathcal{O}(\lambda^6) \; .
\end{eqnarray}
It is clear that $\widetilde{\rho} \simeq \overline{\rho} \simeq \rho$ and
$\widetilde{\eta} \simeq \overline{\eta} \simeq \eta$ hold in the leading-order
approximation, implying a congruence between the rescaled unitarity triangles
$\triangle^\prime_c$ and $\triangle^\prime_s$.
The analytical approximations made in Eqs.~(6) and (7) allow us to see some fine
differences between the apexes $(\overline{\rho} , \overline{\eta})$ and
$(\widetilde{\rho} , \widetilde{\eta})$ of these two triangles as follows:
\begin{eqnarray}
\widetilde{\rho} - \overline{\rho} \simeq \left[\rho \left( 1 - \rho\right)
+ \eta^2 \right] \lambda^2 \; , \quad
\widetilde{\eta} - \overline{\eta} \simeq \eta \left( 1 - 2\rho \right)
\lambda^2 \; ,
\end{eqnarray}
up to the accuracy of ${\cal O}(\lambda^4)$.

Given the values of $\lambda$, $A$, $\overline{\rho}$ and $\overline{\eta}$ which
have been determined from a global analysis of current experimental data on quark
flavor mixing and CP violation \cite{Tanabashi:2018oca},
\begin{eqnarray}
\lambda = 0.22453 \pm 0.00044 \; , \quad
A = 0.836 \pm 0.015 \; , \quad
\overline{\rho} = 0.122^{+0.018}_{-0.017} \; , \quad
\overline{\eta} = 0.355^{+0.012}_{-0.011} \; ,
\end{eqnarray}
we immediately obtain
\begin{eqnarray}
\rho = 0.125 \pm 0.018 \; , \quad \eta = 0.364 \pm 0.012
\end{eqnarray}
from the exact relationship between $(\rho, \eta)$ and
$(\overline{\rho} , \overline{\eta})$ \cite{Buras:1994ec,Charles:2004jd,
Tanabashi:2018oca}:
\begin{eqnarray}
\rho + {\rm i} \eta = \frac{ \sqrt{1-A^2\lambda^4} \left(\overline{\rho} +
{\rm i} \overline{\eta}\right)}{\sqrt{1-\lambda^2} \left[ 1- A^2\lambda^4
\left(\overline{\rho} + {\rm i} \overline{\eta}\right)\right]} \; .
\end{eqnarray}
If the apex of the rescaled unitarity triangle $\triangle^\prime_c$ can be
experimentally fixed (i.e., if $\widetilde{\rho}$ and $\widetilde{\eta}$ can
be directly determined from some precision measurements), it will also be
possible to figure out the values of $\rho$ and $\eta$ from the exact
relationship between $(\rho, \eta)$ and $(\widetilde{\rho} , \widetilde{\eta})$:
\begin{eqnarray}
\rho + {\rm i} \eta = \frac{ \sqrt{1-\lambda^2} \left(\widetilde{\rho} +
{\rm i} \widetilde{\eta}\right)}{\sqrt{1-A^2\lambda^4} \left[ 1- \lambda^2
\left(\widetilde{\rho} + {\rm i} \widetilde{\eta}\right)\right]} \; .
\end{eqnarray}
Then a comparison between the results of $\rho$ and $\eta$ obtained from
Eqs.~(11) and (12) will provide a meaningful consistency check of the CKM
picture for CP violation described by the twin $b$-flavored unitarity triangles
$\triangle^\prime_s$ and $\triangle^\prime_c$ within the SM.

Unfortunately, so far no effort has been made towards establishing
$\triangle^\prime_c$ from the available experimental data.
It is well known that the apex of $\triangle^\prime_s$ has
been excessively constrained by current experimental results for
$|V^{}_{ub}|/|V^{}_{cb}|$, $\sin 2\beta$ (CP violation in $B^0_d$ vs
$\bar{B}^0_d \to J/\psi K^{}_{\rm S}$ decays), $\Delta m^{}_d$ (the mass
difference of $B^0_d$-$\bar{B}^0_d$ mixing), $\Delta m^{}_s$ (the mass
difference of $B^0_s$-$\bar{B}^0_s$ mixing), $\varepsilon^{}_K$ (CP violation
in $K^0$-$\bar{K}^0$ mixing) and so on
\cite{Charles:2004jd,Tanabashi:2018oca,Bona:2005vz}. To separately constrain
the apexes of $\triangle^\prime_s$ and $\triangle^\prime_c$, one may make use
of the experimental data from $B^{\pm}_u$ and $B^0_d$-$\bar{B}^0_d$ systems and
those from $B^{\pm}_u$ and $B^0_s$-$\bar{B}^0_s$ systems, respectively.
The measurements of $\Delta m^{}_d$ and $\Delta m^{}_s$, which depend respectively
on $V^*_{tb} V^{}_{td}$ and $V^*_{tb} V^{}_{ts}$ via the $t$-dominated box
diagrams, are expected to be useful to
distinguish between the apexes of $\triangle^\prime_s$ and $\triangle^\prime_c$.
It is also possible to determine $|V^*_{tb} V^{}_{ts}|$ from more precise
measurements of $B \to X^{}_s \gamma$ and $B^{}_s \to \mu^+\mu^-$ decays in
the near future \cite{Tanabashi:2018oca}.

In the present case what we can do is to calculate $\widetilde{\rho}$ and
$\widetilde{\eta}$ by means of of Eq.~(12) with the values of $\rho$ and
$\eta$ obtained in Eq.~(10). As a result,
\begin{eqnarray}
\widetilde{\rho} = 0.134 \pm 0.018 \; , \quad
\widetilde{\eta} = 0.368 \pm 0.012 \; .
\end{eqnarray}
Eqs.~(9) and (13) tell us that the differences $\widetilde{\rho} - \overline{\rho}$
and $\widetilde{\eta} - \overline{\eta}$ remain within the error bars of
these four parameters. This observation means that the twin $b$-flavored triangles
$\triangle^\prime_c$ and $\triangle^\prime_s$ will be experimentally
distinguishable at the $3\sigma$ level if their apexes
$(\overline{\rho} , \overline{\eta})$ and
$(\widetilde{\rho} , \widetilde{\eta})$ can be determined to the $0.4\%$
precision. This is certainly a big challenge.

Let us proceed to take a look at the inner angles of triangles $\triangle^\prime_s$
and $\triangle^\prime_c$. Up to the accuracy of ${\cal O}(\lambda^6)$, we obtain
\begin{eqnarray}
\alpha & \equiv & \arg \left( - \frac{V^{*}_{tb} V^{}_{td}}{V^{*}_{ub}V^{}_{ud}}
\right) = \arg \left(- \frac{1-\overline{\rho} - {\rm i} \overline{\eta}}{\bar{\rho}
+ {\rm i} \overline{\eta}}\right) =
\arctan\left(\frac{\eta}{\eta^2 + \rho^2 - \rho} \right) + \frac{\eta}
{\displaystyle 2\left[\eta^2 + \left(\rho-1\right)^2\right]} \lambda^2
\nonumber \\
& & + \left\{ \frac{\eta \left(4A^2+3\right)}{\displaystyle
8\left[\eta^2 +(\rho-1)^2\right]} - \frac{\eta \left(1-\rho\right)}
{\displaystyle 4\left[\eta^2 +(\rho-1)^2\right]^2} \right\} \lambda^4
+ \mathcal{O}(\lambda^6) \; ,
\nonumber \\
\beta & \equiv & \arg \left( -\frac{V^{*}_{cb}V^{}_{cd}}{V^{*}_{tb}V^{}_{td}}
\right) = \arg \left(\frac{1}{1-\overline{\rho} - {\rm i} \overline{\eta}}\right)
= \arctan \left( \frac{\eta}{1-\rho} \right) - \frac{\eta}{2\left[\eta^2
+ \left(\rho-1\right)^2\right]} \lambda^2
\nonumber \\
& & + \left\{ \eta A^2 - \frac{\eta\left(4A^2+3\right)}{8\left[\eta^2 +
\left(\rho-1\right)^2\right]} + \frac{\eta\left(1-\rho\right)}{4\left[\eta^2
+ \left(\rho-1\right)^2\right]^2} \right\} \lambda^4 + \mathcal{O}(\lambda^6) \; ,
\nonumber \\
\gamma & \equiv & \arg \left( - \frac{V^{*}_{ub}V^{}_{ud}}{V^{*}_{cb}V^{}_{cd}}
\right) = \arg \left(\overline{\rho} + {\rm i} \overline{\eta}\right)
= \arctan \left(\frac{\eta}{\rho}\right) - \eta A^2\lambda^4 +
\mathcal{O}(\lambda^6) \; ;
\end{eqnarray}
and
\begin{eqnarray}
\alpha^\prime & \equiv & \arg \left( - \frac{V^{*}_{ud} V^{}_{td}}{V^{*}_{ub}
V^{}_{tb}} \right) = \arg \left(- \frac{1-\widetilde{\rho} - {\rm i}
\widetilde{\eta}}{\widetilde{\rho} + {\rm i}\widetilde{\eta}} \right)
= \arctan\left(\frac{\eta}{\eta^2 + \rho^2 - \rho} \right) + \frac{\eta}
{\displaystyle 2\left[\eta^2 + \left(\rho-1\right)^2\right]} \lambda^2
\nonumber \\
& & + \left\{ \frac{\eta \left(4A^2+3\right)}{\displaystyle
8\left[\eta^2 +(\rho-1)^2\right]} - \frac{\eta \left(1-\rho\right)}
{\displaystyle 4\left[\eta^2 +(\rho-1)^2\right]^2} \right\} \lambda^4
+ \mathcal{O}(\lambda^6) \; ,
\nonumber \\
\beta^\prime & \equiv & \arg \left( - \frac{V^{*}_{us}V^{}_{ts}}{V^{*}_{ud}V^{}_{td}}
\right) = \arg \left(\frac{1}{1-\widetilde{\rho} - {\rm i}\widetilde{\eta}}\right)
= \arctan \left( \frac{\eta}{1-\rho} \right) + \left\{ \eta -\frac{\eta}
{2\left[\eta^2 + \left(\rho-1\right)^2\right]} \right\}\lambda^2
\nonumber \\
& & - \left\{  \frac{\eta\left(4A^2+3\right)}{8\left[\eta^2 + \left(\rho-1\right)^2
\right]} - \frac{\eta\left(1-\rho\right)}{4\left[\eta^2 + \left(\rho-1\right)^2
\right]^2} - \eta \left(\frac{1}{2}-\rho\right)\right\} \lambda^4
+ \mathcal{O}(\lambda^6) \; ,
\nonumber \\
\gamma^\prime & \equiv & \arg \left( - \frac{V^{*}_{ub}V^{}_{tb}}{V^{*}_{us}V^{}_{ts}}
\right) = \arg \left(\widetilde{\rho} + {\rm i} \widetilde{\eta}\right)
= \arctan \left(\frac{\eta}{\rho}\right) - \eta \lambda^2 - \eta
\left(\frac{1}{2} - \rho\right)\lambda^4 + \mathcal{O}(\lambda^6) \; .
\end{eqnarray}
One can see that $\alpha^\prime = \alpha$ holds exactly, and
\begin{eqnarray}
\beta^\prime - \beta = \gamma - \gamma^\prime \simeq \eta \lambda^2 \left[
1 + \left(\frac{1}{2} -A^2 -\rho\right) \lambda^2\right] \; .
\end{eqnarray}
So we arrive at $\beta^\prime > \beta$ and $\gamma > \gamma^\prime$, and the
difference between each pair of these inner angles is measured by $\eta \lambda^2$
in the leading-order approximation. Taking account of the values of
$\lambda$, $A$, $\overline{\rho}$ and $\overline{\eta}$ in Eq.~(9), we immediately
obtain $\alpha = \alpha^\prime \simeq 87.0^\circ \pm 2.5^\circ$,
$\beta \simeq 22.0^\circ \pm 0.8^\circ$,
$\gamma \simeq 71.0^\circ \pm 2.6^\circ$,
$\beta^\prime \simeq 23.0^\circ \pm 0.8^\circ$ and
$\gamma^\prime \simeq 70.0^\circ \pm 2.6^\circ$. So the numerical results of
$\beta^\prime - \beta$ and $\gamma - \gamma^\prime$, which also
characterize the tiny difference between the triangles $\triangle^\prime_s$
and $\triangle^\prime_c$, remain within the error bars of these four angles
and hence require more accurate measurements.

Note that the result obtained in Eq.~(16) is essentially identical to the
smallest inner angle of a ``squashed" CKM unitarity triangle defined by
the relation $V^*_{ub} V^{}_{us} + V^*_{cb} V^{}_{cs} + V^*_{tb} V^{}_{ts} = 0$ in the
complex plane:
\begin{eqnarray}
\beta^{}_s \equiv \arg\left(-\frac{V^*_{tb} V^{}_{ts}}{V^*_{cb} V^{}_{cs}}
\right) \simeq \eta \lambda^2 \left[1 + \left(\frac{1}{2} - \rho\right)
\lambda^2\right] \; ,
\end{eqnarray}
which characterizes the strength of CP violation in $B^0_s$ vs $\bar{B}^0_s
\to J/\psi \phi$ decays. Current experimental constraint on this CP-violating
quantity is $2\beta^{}_s = 0.021 \pm 0.031$ \cite{Amhis:2016xyh}, still too preliminary
to be confronted with the SM prediction $2\beta^{}_s = 0.037 \pm 0.001$ that can be
obtained from Eq.~(17) with the help of Eqs.~(9) and (10). We conclude that
distinguishing between $\triangle^\prime_s$ and $\triangle^\prime_c$ in a
geometrical way requires the same level of high precision as the direct measurement
of $\beta^{}_s$, and hence these two approaches can be complementary to each other
in testing consistency of the CKM picture of quark flavor mixing and CP violation.
\begin{figure}[t]
\centering
\includegraphics[width = 0.98\linewidth]{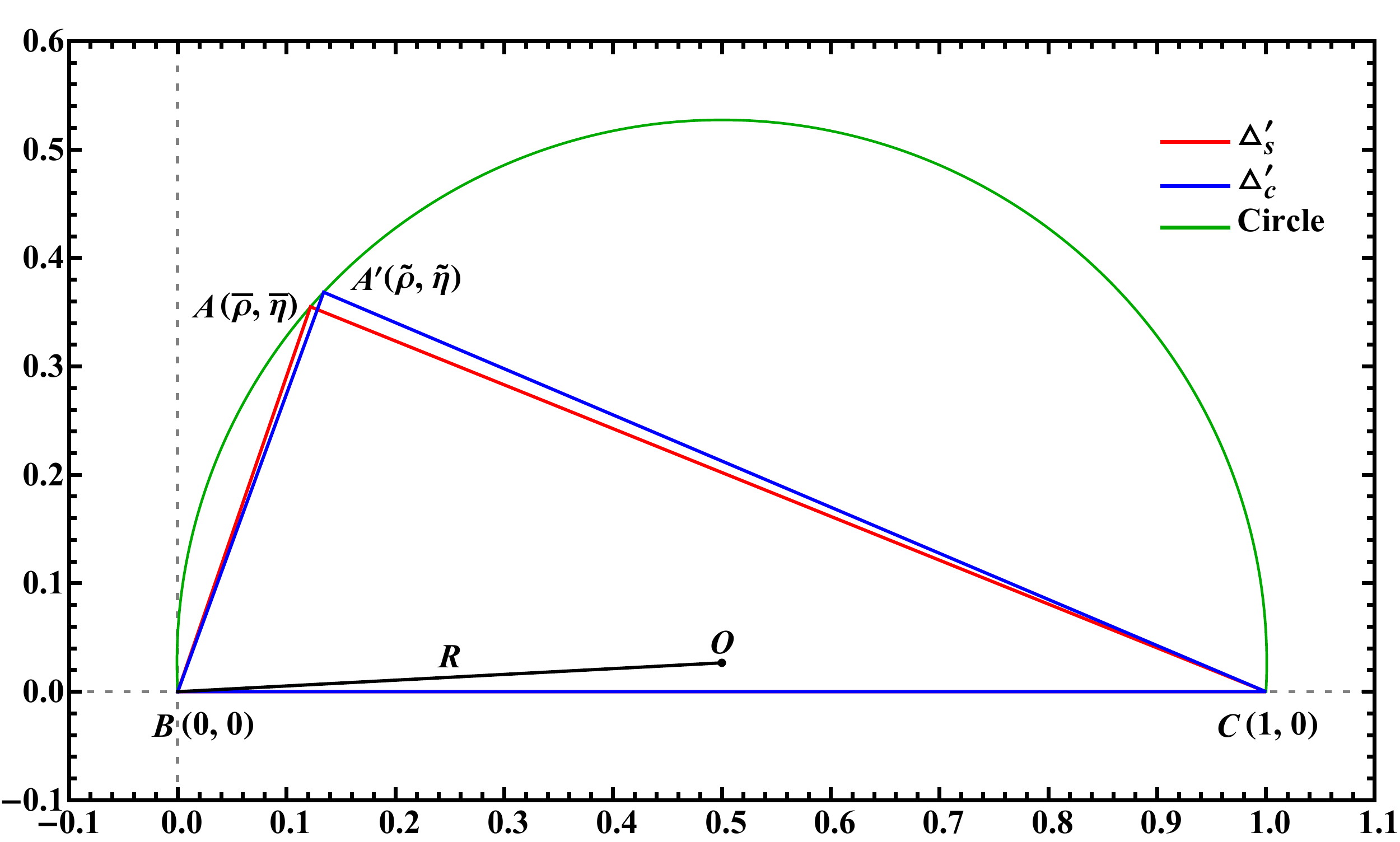}
\caption{The rescaled CKM unitarity triangles $\triangle^\prime_{s} =
\triangle ABC$ and $\triangle^\prime_{c} = \triangle A^\prime BC$ in the complex
plane, and their corresponding inner angles are defined as $\alpha = \angle BAC$,
$\beta = \angle ACB$, $\gamma = \angle ABC$ and $\alpha^\prime = \angle BA^\prime C$,
$\beta^\prime= \angle A^\prime CB$, $\gamma^\prime = \angle A^\prime BC$. Note that
$\alpha^\prime = \alpha$ holds by definition, as shown in Eqs.~(14) and (15).
The apexes $A (\overline{\rho}, \overline{\eta})$ and
$A^\prime (\widetilde{\rho}, \widetilde{\eta})$ of these two triangles are located
on the same circular arc, whose center and radius are $O = (0.5, 0.5 \cot\alpha)$ and
$R = 0.5 \csc\alpha$ respectively.}
\end{figure}

Now that the twin rescaled unitarity triangles $\triangle^\prime_c$
and $\triangle^\prime_s$ share a common inner angle
$\alpha^\prime = \alpha$ and a common side $BC$ as shown in Figure 1, their
corresponding apexes $(\widetilde{\rho},\widetilde{\eta})$ and
$(\overline{\rho},\overline{\eta})$ must be on a circular arc in the upper complex
plane. To see this interesting point in a more transparent way, let us first write out
\begin{eqnarray}
\cos\alpha = \frac{\overline{\rho}^2 + \overline{\eta}^2 +
\left(1 - \overline{\rho}\right)^2 + \overline{\eta}^2 - 1}
{2\sqrt{\overline{\rho}^2 + \overline{\eta}^2} \sqrt{\left(1 - \overline{\rho}
\right)^2 + \overline{\eta}^2}} \;
\end{eqnarray}
for the apex $(\overline{\rho},\overline{\eta})$ with the help of the cosine theorem,
and then express this equation as
\begin{eqnarray}
\left(\overline{\rho} - \frac{1}{2}\right)^2 + \left(\overline{\eta} - \frac{1}{2}\cot{\alpha}\right)^2 = \left(\frac{1}{2} \csc\alpha \right)^2 \; .
\end{eqnarray}
It becomes obvious that Eq.~(18) or (19) actually defines a circular arc in the
upper complex plane, whose center and radius are
$O = \left(0.5, 0.5\cot\alpha\right)$ and $R = 0.5\csc\alpha$
respectively. Because Eqs.~(18) and (19) keep unchanged with the replacements
$\overline{\rho} \to \widetilde{\rho}$ and $\overline{\eta} \to \widetilde{\eta}$,
the apex $(\widetilde{\rho},\widetilde{\eta})$ must be located on the same
circular arc. This observation geometrically illustrates how similar
$\triangle^\prime_c$ and $\triangle^\prime_s$ are, and it can be extended
to similar discussions about the other two pairs of the CKM unitarity
triangles, whose shapes are nevertheless very sharp \cite{Xing:2019vks,Bigi:1999hr}.

The fact that all the apexes of $\triangle^\prime_c$ and $\triangle^\prime_s$
are located on the same circular arc is of course a natural consequence of unitarity
of the CKM matrix $V$. It provides another interesting way to test the consistency of
quark flavor mixing and CP violation in the SM. Since $\triangle^\prime_s$
has been established to a very good degree of accuracy, it allows us to fix a
benchmark circular arc as shown in Figure~1. The future measurements of
$(\widetilde{\rho},\widetilde{\eta})$ will tell us to what extent the experimental
values of this apex are also located on the same circular arc. In other words,
an experimental test of the equality
\begin{eqnarray}
\left(\widetilde{\rho} - \frac{1}{2}\right)^2 + \left(\widetilde{\eta} - \frac{1}{2}\cot{\alpha}\right)^2 =
\left(\overline{\rho} - \frac{1}{2}\right)^2 + \left(\overline{\eta} - \frac{1}{2}\cot{\alpha}\right)^2
\end{eqnarray}
will be greatly useful at both the super-$B$ factory and the high-luminosity LHC.
It is worth mentioning that the possibility of $\alpha = \pi/2$, which is
compatible with the experimental result $\alpha = \left(84.5^{+5.9}_{-5.2}\right)^\circ$
extracted from $B \to \pi\pi$, $\rho\pi$ and $\rho\rho$ decay modes
\cite{Charles:2004jd,Tanabashi:2018oca},
has been conjectured in exploring realistic textures of quark mass matrices or
studying phenomenological implications of the CKM matrix (e.g.,
Refs. \cite{Fritzsch:1995nx,Fritzsch:1999rb,Fritzsch:2002ga,Koide:2004gj,Masina:2006ad,
Harrison:2007yn,Xing:2009eg,Antusch:2009hq}). In this special but suggestive case
$\triangle^\prime_c$ and $\triangle^\prime_s$ are exactly the right triangles,
from which one is left with $O = (0.5, 0)$ and $R = 0.5$ for the circular arc
in Figure~1.

\section{Two-loop RGE evolution of $\triangle^\prime_s$ and $\triangle^\prime_c$}

Note that elements of the CKM matrix $V$ depend on the energy scale $\Lambda$,
but they are usually treated as constants below $\Lambda = M^{}_W$.
When $\Lambda$ is far above the electroweak scale
$\Lambda^{}_{\rm EW} \sim 10^2~{\rm GeV}$, the RGE running effects of
$V$ will become appreciable and should be taken into account
(see Ref.~\cite{Xing:2019vks} for a recent review). In particular, the two-loop RGEs
of $V$ have been derived for $\Lambda$ evolving between $\Lambda^{}_{\rm EW}$ and
$\Lambda^{}_{\rm GUT} \sim 10^{16} ~{\rm GeV}$ in the framework of
the SM or its minimal supersymmetric version (MSSM)
\cite{Machacek:1983fi,Barger:1992ac,Barger:1992pk,Luo:2002ey}. In view of the
very strong hierarchies of quark Yukawa couplings of the same electric charge (i.e.,
$y^{}_u/y^{}_c \sim y^{}_c/y^{}_t \sim \lambda^4$ and $y^{}_d/y^{}_s \sim
y^{}_s/y^{}_b \sim \lambda^2$ at a given energy scale \cite{Xing:2007fb,Xing:2011aa})
and the relatively strong hierarchies of those off-diagonal elements of $V$,
Barger {\it et al} have
found \cite{Barger:1992pk}
\begin{eqnarray}
&& \frac{\rm d}{{\rm d}t} \left( \begin{matrix} |V^{}_{ud}| & |V^{}_{us}|
& |V^{}_{ub}| \\ |V^{}_{cd}| & |V^{}_{cs}| & |V^{}_{cb}| \\ |V^{}_{td}|
& |V^{}_{ts}| & |V^{}_{tb}| \end{matrix}\right) \simeq
\left(S^{}_{1} + S^{}_{2} \right)  \left( \begin{matrix} 0 & 0 & |V^{}_{ub}| \\ 0
& 0 & |V^{}_{cb}| \\ |V^{}_{td}| & |V^{}_{ts}| & 0 \end{matrix}\right) \; ,
\nonumber \\
&& \frac{{\rm d} {\cal J}}{{\rm d}t} \simeq 2 \left(S^{}_{1} + S^{}_{2} \right)
\mathcal{J} \; ,
\end{eqnarray}
where $t \equiv \ln \left(\Lambda/\Lambda^{}_{\rm EW}\right)$,
${\cal J} \equiv {\rm Im} \left(V^{}_{ud} V^{}_{cs} V^*_{us} V^*_{cd}\right)
= {\rm Im}\left(V^{}_{us} V^{}_{cb} V^*_{ub} V^*_{cs}\right) =
{\rm Im}\left(V^{}_{td} V^{}_{us} V^*_{ts} V^*_{ud}\right) = \cdots$
is the well-known Jarlskog invariant of CP violation
\cite{Jarlskog:1985ht} and ${\cal J} \simeq A^2 \lambda^6 \eta$ holds up to
the accuracy of ${\cal O}(\lambda^8)$ in the Wolfenstein parametrization of
$V$, $S^{}_{1}$ and $S^{}_{2}$ stand respectively for the one- and two-loop
contributions to the RGEs of $V$. To be explicit,
\begin{eqnarray}
S^{}_{1} & = & -\frac{1}{16\pi^2} \left( C^{\rm u}_{\rm d}y^2_t +
C^{\rm d}_{\rm u}y^2_b\right) \; ,
\nonumber \\
S^{}_{2} & = & -\frac{1}{\left(16\pi^2\right)^2} \left[ D^{\rm u}_{\rm d}y^2_t
+ D^{\rm d}_{\rm u }y^2_b + \left(D^{\rm ud}_{\rm d} + D^{\rm du}_{\rm u}\right)
y^2_t y^2_b + D^{\rm uu}_{\rm d}y^4_t + D^{\rm dd}_{\rm u} y^4_b \right] \; ,
\end{eqnarray}
where $y^{}_t$ and $y^{}_b$ denote the Yukawa couplings of top and bottom
quarks respectively, and the relevant coefficients on the right-hand side of
Eq.~(22) are
\begin{eqnarray}
&& C^{\rm u}_{\rm d} = C^{\rm d}_{\rm u} = -\frac{3}{2} \; ,
\quad
D^{\rm u}_{\rm d} \simeq -\frac{79}{80} g^2_1 + \frac{9}{16} g^2_2
- 16 g^2_3 + \frac{15}{4} \left(y^2_t + y^2_b\right) \; ,
\nonumber \\
&& D^{\rm d}_{\rm u} \simeq -\frac{43}{80} g^2_1 + \frac{9}{16} g^2_2
- 16 g^2_3 + \frac{15}{4} \left(y^2_t + y^2_b\right) \; ,
\quad
D^{\rm uu}_{\rm d} = D^{\rm dd}_{\rm u} = \frac{11}{4} \; ,
\quad
D^{\rm ud}_{\rm d} = D^{\rm du}_{\rm u} = -1 \; , \hspace{0.6cm}
\end{eqnarray}
in the SM
\footnote{Note that the errors associated with $D^{\rm u}_{\rm d}$ and $D^{\rm d}_{\rm u}$
in the SM case in Refs.~\cite{Machacek:1983fi,Barger:1992pk} were corrected in
Ref.~\cite{Luo:2002ey}.};
or
\begin{eqnarray}
&& C^{\rm u}_{\rm d} = C^{\rm d}_{\rm u} = 1 \; ,
\quad
D^{\rm u}_{\rm d} \simeq \frac{4}{5} g^2_1 - 3 y^2_t \; ,
\quad
D^{\rm d}_{\rm u} \simeq \frac{2}{5} g^2_1 - 3 y^2_b \; , \hspace{0.6cm}
\nonumber \\
&& D^{\rm uu}_{\rm d} = D^{\rm dd}_{\rm u} = -2 \; ,
\quad
D^{\rm ud}_{\rm d} = D^{\rm du}_{\rm u} = 0 \; ,
\end{eqnarray}
in the MSSM with $g^{}_i$ (for $i=1,2,3$) being the respective gauge couplings of
electromagnetic, weak and strong interactions.
After a careful check of the approximations made in obtaining
Eq.~(21), we conclude that the two-loop RGEs shown in Eq.~(21) are actually valid
up to the accuracy of ${\cal O}(\lambda^4)$. To the same order, the two-loop
RGEs of the Wolfenstein parameters can be figured out as follows:
\begin{eqnarray}
\frac{{\rm d} \lambda}{{\rm d} t} \simeq \frac{{\rm d} \rho}{{\rm d} t}
\simeq \frac{{\rm d} \eta}{{\rm d} t} \simeq 0 \; , \quad
\frac{{\rm d} A}{{\rm d} t} \simeq ( S^{}_1 + S^{}_2 ) A \; , \quad
\frac{{\rm d} \overline{\rho}}{{\rm d} t} \simeq \frac{{\rm d} \overline{\eta}}{{\rm d} t}
\simeq \frac{{\rm d} \widetilde{\rho}}{{\rm d} t} \simeq
\frac{{\rm d} \widetilde{\eta}}{{\rm d} t} \simeq 0 \; .
\end{eqnarray}
Two immediate comments are in order.
\begin{itemize}
\item     The rescaled unitarity triangles $\triangle^\prime_s$ and
$\triangle^\prime_c$ keep unchanged when the energy scale $\Lambda$ evolves from
$\Lambda^{}_{\rm EW}$ to $\Lambda^{}_{\rm GUT}$ or vice versa, in a very
good approximation up to the accuracy of ${\cal O}(\lambda^4)$. Accordingly,
three sides of the original unitarity triangle $\triangle^{}_s$ or $\triangle^{}_c$
are rescaled with the same amount as $\Lambda$ evolves, and thus the overall shape
of either of the two triangles keeps undeformed up to the same accuracy. Although
a similar observation has been made with the help of the one-loop RGEs
\cite{Luo:2009wa}, the relevant accuracy is certainly worse than the present
two-loop result.

\item     Given $g^{}_1 \simeq 0.46$, $g^{}_2 \simeq 0.65$ and $g^{}_3 \sim 1.21$
together with $y^{}_t \simeq 1$ and $y^{}_b \simeq y^{}_t/60$ at
$\Lambda = M^{}_Z$ in the SM \cite{Tanabashi:2018oca,Xing:2007fb,Xing:2011aa}, one may
make a rough but instructive estimate
\begin{eqnarray}
\frac{S^{}_2}{S^{}_1} \sim \frac{1}{16 \pi^2} \cdot
\frac{D^{\rm u}_{\rm d} + D^{\rm uu}_{\rm d} y^2_t}{C^{\rm u}_{\rm d}}
\sim {\cal O}(\lambda^2) \; . \hspace{0.6cm}
\end{eqnarray}
It is more difficult to analytically estimate the order of $S^{}_2/S^{}_1$ in the
MSSM, because $y^{}_t$ and $y^{}_b$ depend on $\tan\beta$ in a different way. In
Figure 2 we illustrate the magnitudes of $S^{}_1$, $S^{}_2$ and $S^{}_2/S^{}_1$
changing with the energy scale $\Lambda$, where $M^{}_H \simeq 125 ~{\rm GeV}$ in
the SM and $\tan\beta \simeq 10$ or $30$ in the MSSM are input.
It becomes clear that the two-loop effect is suppressed by a factor of
${\cal O}(\lambda^3)$ to ${\cal O}(\lambda^2)$ in magnitude as compared
with the dominant one-loop contribution, but it definitely makes sense for
us to keep the former in the approximations made in Eq.~(21).
\end{itemize}
Moreover, it is well known that the area of each of the six CKM unitarity triangles
equals ${\cal J}/2$, and hence the two-loop RGE evolution of $\cal J$ is consistent with
that of $A^2$ in the Wolfenstein parametrization of $V$ up to the accuracy of
${\cal O} (\lambda^4)$.
\begin{figure}[h]
	\centering
	\includegraphics[width = 0.98\linewidth]{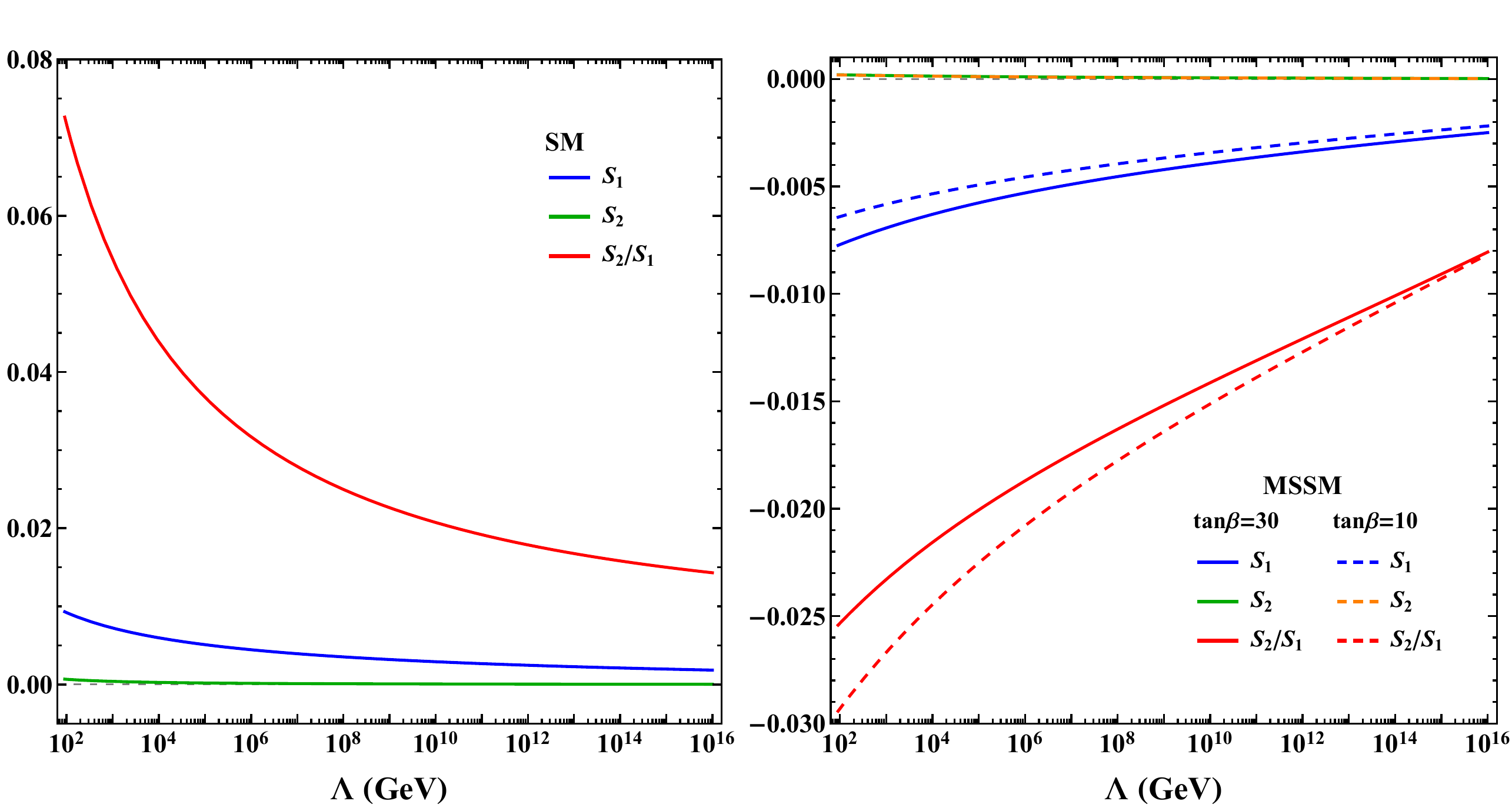}
    \vspace{0cm}
	\caption{A comparison between one- and two-loop contributions to the RGEs
in Eq.~(22) in the SM with $M^{}_H \simeq 125 ~{\rm GeV}$ (left panel) and the MSSM with
$\tan\beta =10$ or $30$ (right panel).}
\end{figure}

The integral form of Eq.~(25) can be obtained in a straightforward way as follows:
\begin{eqnarray}
&& \lambda(\Lambda) \simeq \lambda(\Lambda_{\rm EW}) \;, \quad
\rho(\Lambda) \simeq \rho(\Lambda_{\rm EW}) \;, \quad
\eta(\Lambda) \simeq \eta(\Lambda_{\rm EW}) \;, \quad
A(\Lambda) \simeq I^{}_{1} I^{}_{2} A(\Lambda_{\rm EW}) \; ; \hspace{0.6cm}
\nonumber \\
&& \overline{\rho}(\Lambda) \simeq \overline{\rho}(\Lambda_{\rm EW}) \;, \quad
\overline{\eta}(\Lambda) \simeq \overline{\eta}(\Lambda_{\rm EW}) \;, \quad
\widetilde{\rho}(\Lambda) \simeq \widetilde{\rho}(\Lambda_{\rm EW}) \;, \quad
\widetilde{\eta}(\Lambda) \simeq \widetilde{\eta}(\Lambda_{\rm EW}) \;,
\end{eqnarray}
where $\Lambda$ denotes an arbitrary energy scale between $\Lambda^{}_{\rm EW}$
and $\Lambda^{}_{\rm GUT}$, and the loop functions $I^{}_i$ (for $i=1,2$)
are simply defined as
\begin{eqnarray}
I^{}_{i} &=& \exp \left(\int^{\ln(\Lambda/\Lambda_{\rm EW})}_{0} S^{}_{i} {\rm d}t
\right) \; .
\end{eqnarray}
Of course, $|V^{}_{ub}|$, $|V^{}_{cb}|$, $|V^{}_{td}|$ and $|V^{}_{ts}|$ evolve
in the same way as $A$, while
$\mathcal{J}(\Lambda) \simeq I^2_{1} I^2_{2} \mathcal{J}(\Lambda_{\rm EW})$ holds
for the Jarlskog invariant. In comparison, $|V^{}_{ud}|$, $|V^{}_{us}|$, $|V^{}_{cd}|$,
$|V^{}_{cs}|$ and $|V^{}_{tb}|$ are essentially stable against changes of the energy
scale $\Lambda$.

For the sake of illustration, we show the running behaviors of $I^{}_1$, $I^{}_2$
and $I^{}_1 I^{}_2$ in Figure~3, where $M^{}_H \simeq 125~{\rm GeV}$ has been
input for the SM and $\tan\beta = 10$ or $30$ has been taken for the MSSM. One
can see that the overall RGE-induced scaling effect on $A$ from
$\Lambda^{}_{\rm EW}$ and $\Lambda^{}_{\rm GUT}$ or vice versa is about $10\%$
for the inputs under consideration.
\begin{figure}[h]
	\centering
	\includegraphics[width = 0.98\linewidth]{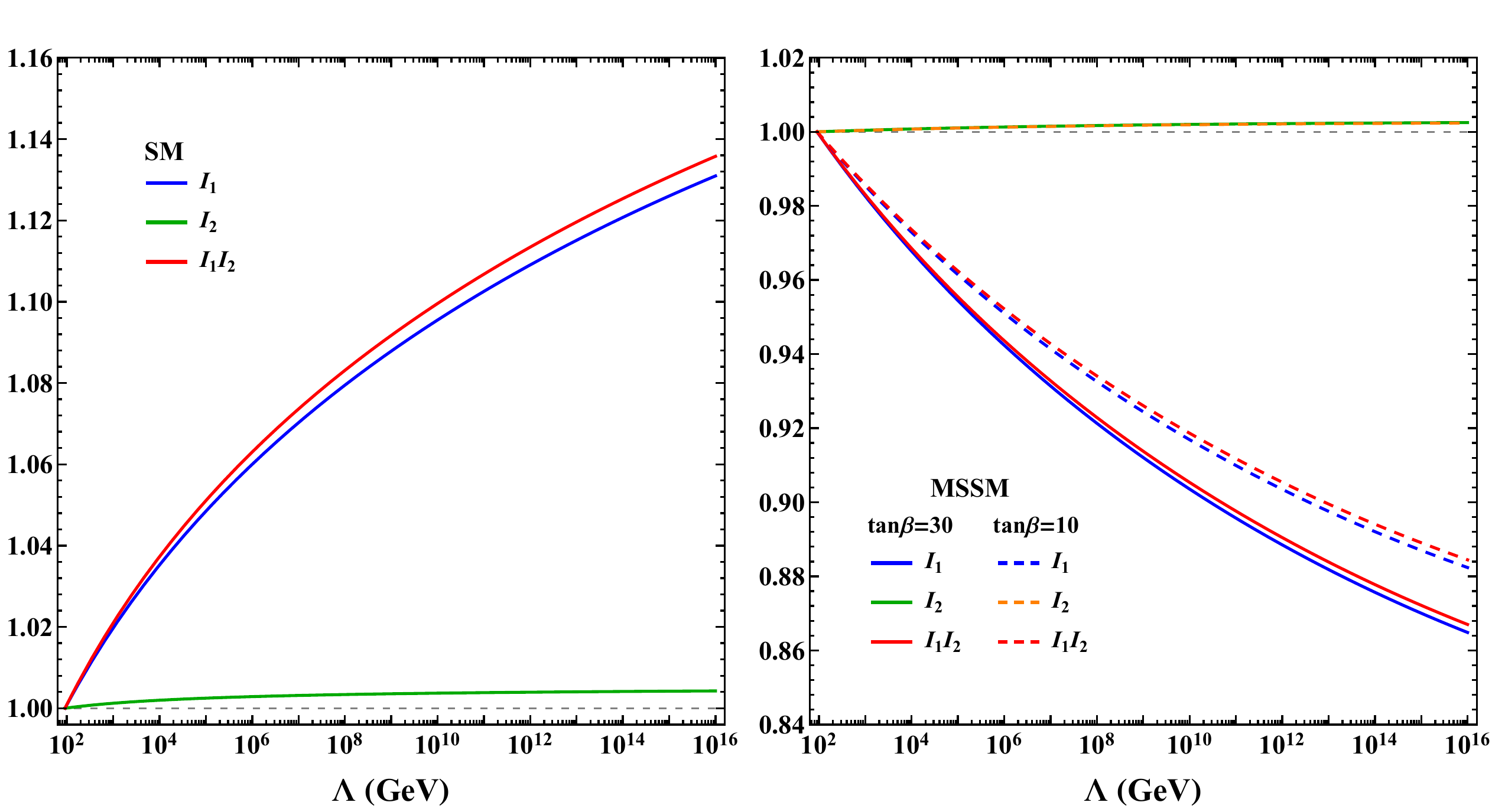}
    \vspace{0cm}
	\caption{The two-loop RGE evolution functions $I^{}_1$, $I^{}_2$ and $I^{}_1 I^{}_2$
in the SM with $M^{}_H \simeq 125 ~{\rm GeV}$ (left panel) and the MSSM with
$\tan\beta =10$ or $30$ (right panel).}
\end{figure}

\section{Comments on leptonic unitarity triangles}

In the lepton sector, the $3\times 3$ Pontecorvo-Maki-Nakagawa-Sakata (PMNS)
flavor mixing matrix $U$ \cite{Maki:1962mu,Pontecorvo:1967fh} can
geometrically be described by three pairs of unitarity triangles in the complex plane
\cite{Fritzsch:1999ee}
\footnote{Note that whether the PMNS matrix $U$ is exactly unitary or not depends
the mechanism of neutrino mass generation, but its possible unitarity-violating effects
must be less than one percent as constrained by current precision electroweak data and
neutrino oscillation data \cite{Xing:2019vks}.}.
Here we focus on the rescaled versions of these triangles, defined as
\begin{eqnarray}
\triangle^\prime_e &:& \;\; 1 + \frac{U^{}_{\mu2}U^{\ast}_{\tau2}}{U^{}_{\mu1}U^\ast_{\tau1}} +
\frac{U^{}_{\mu3}U^{\ast}_{\tau3}}{U^{}_{\mu1}U^\ast_{\tau1}} = 0 \;,
\nonumber
\\
\triangle^\prime_1 &:& \;\; 1 + \frac{U^{}_{\mu3}U^{\ast}_{\mu2}}{U^{}_{e3}U^\ast_{e2}} +
\frac{U^{}_{\tau3}U^{\ast}_{\tau2}}{U^{}_{e3}U^\ast_{e2}} = 0 \;,
\end{eqnarray}
\begin{eqnarray}
\triangle^\prime_\mu &:& \;\; 1 + \frac{U^{}_{\tau1}U^{\ast}_{e1}}{U^{}_{\tau2}U^\ast_{e2}} +
\frac{U^{}_{\tau3}U^{\ast}_{e3}}{U^{}_{\tau2}U^\ast_{e2}} = 0 \;,
\nonumber
\\
\triangle^\prime_2 &:& \;\; 1 + \frac{U^{}_{e1}U^{\ast}_{e3}}{U^{}_{\mu1}U^\ast_{\mu3}} +
\frac{U^{}_{\tau1}U^{\ast}_{\tau3}}{U^{}_{\mu1}U^\ast_{\mu3}} = 0 \;,
\end{eqnarray}
and
\begin{eqnarray}
\triangle^\prime_\tau &:& \;\; 1 + \frac{U^{}_{e1}U^{\ast}_{\mu1}}{U^{}_{e3}U^\ast_{\mu3}} +
\frac{U^{}_{e2}U^{\ast}_{\mu2}}{U^{}_{e3}U^\ast_{\mu3}} = 0 \;,
\nonumber
\\
\triangle^\prime_3 &:& \;\; 1 + \frac{U^{}_{e2}U^{\ast}_{e1}}{U^{}_{\tau2}U^\ast_{\tau1}} +
\frac{U^{}_{\mu2}U^{\ast}_{\mu1}}{U^{}_{\tau2}U^\ast_{\tau1}} = 0 \;,
\end{eqnarray}
where $U_{\alpha i}$ ($\alpha = e, \mu, \tau$ and $i = 1, 2, 3$) are the elements of the
PMNS matrix. Note that these six rescaled triangles are completely insensitive to the
unknown Majorana phases of $U$. Note also that each pair of the PMNS unitarity triangles
share a common inner angle, which can be defined as follows
\footnote{In defining $\triangle^{\prime}_s$ and
$\triangle^{\prime}_c$ in the quark sector, we have chosen the longest side of
the unitarity triangle $\triangle^{}_s$ or $\triangle^{}_c$
to rescale its two shorter sides, as shown in Figure~1. Since
current experimental data on lepton flavor mixing and CP violation involve much
larger uncertainties, it is difficult for us to choose a suitable side of a given
PMNS unitarity triangle to rescale the other two sides. So the discussions here
are mainly for illustration.}:
\begin{eqnarray}
\alpha^{}_e \equiv \arg\left(-\frac{U^{}_{\mu 2} U^*_{\tau 2}}{U^{}_{\mu 3}
U^*_{\tau 3}}\right) = \arg\left(-\frac{U^{}_{\tau 3} U^*_{\tau 2}}{U^{}_{\mu 3}
U^*_{\mu 2}}\right) \equiv \alpha^{}_1 \; ,
\nonumber \\
\alpha^{}_\mu \equiv \arg\left(-\frac{U^{}_{\tau 3} U^*_{e 3}}{U^{}_{\tau 1}
U^*_{e 1}}\right) = \arg\left(-\frac{U^{}_{e 1} U^*_{e 3}}{U^{}_{\tau 1}
U^*_{\tau 3}}\right) \equiv \alpha^{}_2 \; ,
\nonumber \\
\alpha^{}_\tau \equiv \arg\left(-\frac{U^{}_{e 1} U^*_{\mu 1}}{U^{}_{e 2}
U^*_{\mu 2}}\right) = \arg\left(-\frac{U^{}_{\mu 2} U^*_{\mu 1}}{U^{}_{e 2}
U^*_{e 1}}\right) \equiv \alpha^{}_3 \; ,
\end{eqnarray}
In a way similar to Eqs.~(3) and (4) for the two $b$-flavored unitarity triangles, one may
define apexes of the above six PMNS unitarity triangles in the complex plane:
\begin{eqnarray}
\rho^{}_e + {\rm i} \eta^{}_e &=& - \frac{U^{}_{\mu3}U^{\ast}_{\tau3}}{U^{}_{\mu1}U^\ast_{\tau1}}
\;, \quad
\rho^{}_1 + {\rm i} \eta^{}_1 = - \frac{U^{}_{\mu3}U^{\ast}_{\mu2}}{U^{}_{e3}U^\ast_{e2}} \;,
\nonumber
\\
\rho^{}_\mu + {\rm i} \eta^{}_\mu &=& - \frac{U^{}_{\tau1}U^{\ast}_{e1}}{U^{}_{\tau2}U^\ast_{e2}}
\;, \quad
\rho^{}_2 + {\rm i} \eta^{}_2 = - \frac{U^{}_{\tau1}U^{\ast}_{\tau3}}{U^{}_{\mu1}U^\ast_{\mu3}} \;,
\nonumber
\\
\rho^{}_\tau + {\rm i} \eta^{}_\tau &=& - \frac{U^{}_{e2}U^{\ast}_{\mu2}}{U^{}_{e3}U^\ast_{\mu3}}
\;, \quad
\rho^{}_3 + {\rm i} \eta^{}_3 = - \frac{U^{}_{e2}U^{\ast}_{e1}}{U^{}_{\tau2}U^\ast_{\tau1}} \;.
\end{eqnarray}
Then Eq.~(32) allows us to show that the apexes $(\rho^{}_e, \eta^{}_e)$
and $(\rho^{}_1 , \eta^{}_1)$ must be located on a circular arc, whose center and radius
are given by $O = (0.5, 0.5 \cot\alpha^{}_e)$ and $R = 0.5 \csc\alpha^{}_e$
respectively:
\begin{eqnarray}
\left(\rho^{}_e - \frac{1}{2}\right)^2 + \left(\eta^{}_e - \frac{1}{2}\cot{\alpha^{}_e}\right)^2
= \left(\rho^{}_1 - \frac{1}{2}\right)^2 + \left(\eta^{}_1 - \frac{1}{2}\cot{\alpha^{}_e}\right)^2
= \left(\frac{1}{2} \csc\alpha^{}_e \right)^2 \; .
\end{eqnarray}
Of course, the apexes $(\rho^{}_\mu, \eta^{}_\mu)$ and $(\rho^{}_2 , \eta^{}_2)$
are also on a circular arc defined by
$O = (0.5, 0.5 \cot\alpha^{}_\mu)$ and $R = 0.5 \csc\alpha^{}_\mu$,
and the apexes $(\rho^{}_\tau, \eta^{}_\tau)$ and $(\rho^{}_3 , \eta^{}_3)$
are located on another circular arc with
$O = (0.5, 0.5 \cot\alpha^{}_\tau)$ and $R = 0.5 \csc\alpha^{}_\tau$.
Unlike the CKM matrix $V$ discussed above, the PMNS matrix $U$ does not have a strong
hierarchy in its structure, and hence it is difficult to make an expansion of $U$
in powers of a small parameter \cite{Xing:2002az}.
\begin{figure}[t]
	\centering
	\includegraphics[width = 0.98\linewidth]{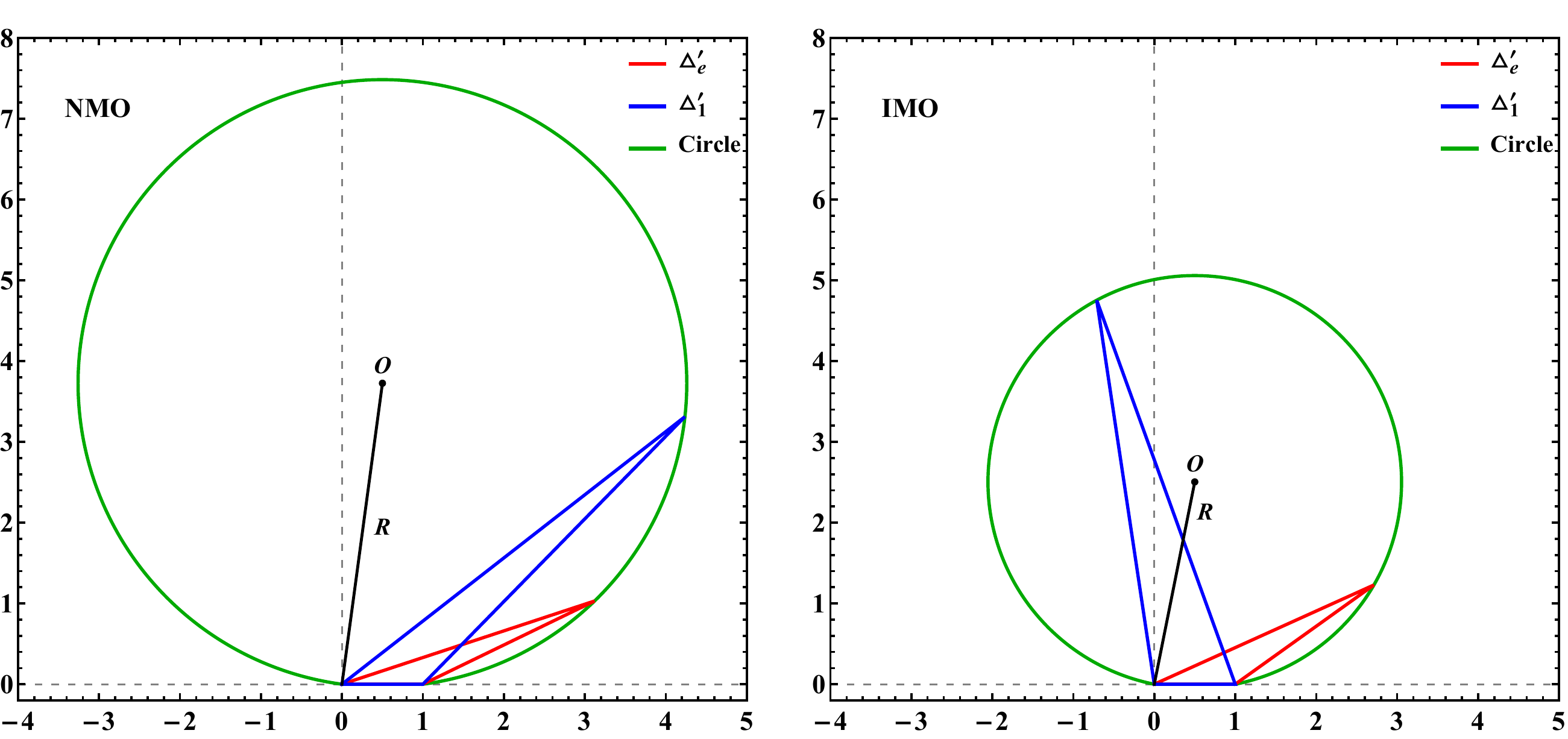}
    \vspace{0cm}
	\caption{The rescaled PMNS unitarity triangles $\triangle^\prime_e$ and $\triangle^\prime_1$
    in the complex plane for either the normal neutrino mass ordering (left panel) or the inverted
    mass ordering (right panel).}
\end{figure}

To illustrate, we plot each pair of the rescaled PMNS unitarity triangles in the
complex plane in Figures~4---6 and calculate not only the center and radius of the
circular arc on which they are located but also their common inner angle in Table~1 by using
the best-fit values of three neutrino mixing angles ($\theta^{}_{12}$, $\theta^{}_{13}$
and $\theta^{}_{23}$) and the Dirac CP-violating phase ($\delta$)~\cite{Capozzi:2018ubv,Esteban:2018azc},
\begin{eqnarray}
\sin^2\theta^{}_{12} = \left\{ \begin{array}{l}
0.310 \\ 0.310 \end{array} \right. \; , \quad
\sin^2\theta^{}_{13} = \left\{ \begin{array}{l}
0.02241 \\ 0.02261 \end{array} \right. \; , \quad
\sin^2\theta^{}_{23} = \left\{ \begin{array}{l}
0.558 \\ 0.563 \end{array} \right. \; , \quad
\delta = \left\{ \begin{array}{l}
222^\circ \\ 285^\circ \end{array} \right. \; ,
\end{eqnarray}
where both the normal neutrino mass ordering (NMO, upper values) and the inverted
mass ordering (IMO, lower values) are taken into account. We admit that for the time
being the large uncertainty associated with $\delta$ prevents us
from establishing the leptonic unitarity triangles in a reliable way
\cite{Gonzalez-Garcia:2014bfa,Xing:2015wzz}, but Figures~4---6 and Table~1 do give one a
ball-park feeling of some salient features of unitarity triangles in the lepton sector. In
particular, it seems much easier for us to distinguish between each pair of the rescaled
PMNS unitarity triangles on a circular arc, simply because their shapes and apexes
are quite different. But this observation might change once more precise data are available.
\begin{table}
\centering
\caption{Numerical results for the center $O$ and radius $R$ of a circular arc on which
each pair of the PMNS unitarity triangles are located, and those for their common inner angle
$\alpha^{}_i$ (for $i=1,2,3$) defined in Eq.~(32) in the NMO or IMO case, where the
best-fit values of three neutrino mixing angles and
the Dirac CP-violating phase \cite{Capozzi:2018ubv,Esteban:2018azc} have been input.}
\vspace{0.3cm}
\begin{tabular}{ccccccc}
    \hline\hline
    \multirow{2}*{Triangles} & \multicolumn{3}{c}{Normal mass ordering (NMO) } & \multicolumn{3}{c}{Inverted mass ordering (IMO) } \\
    \cline{2-4}\cline{5-7}
    & Center $O$ & Radius $R$ & Inner angle $\alpha^{}_i$ & Center $O$ & Radius $R$ & Inner angle $\alpha^{}_i$  \\
    \hline
    $\triangle^{\prime}_{e,1}$ & $(0.5,3.73)$ & $3.76$ & $7.64^\circ$ & $(0.5,2.50)$ & $2.55$ & $11.29^\circ$ \\
    $\triangle^{\prime}_{\mu,2}$ & $(0.5,0.70)$ & $0.86$ & $35.39^\circ$ & $(0.5,-0.03)$ & $0.50$ & $93.62^\circ$\\
    $\triangle^{\prime}_{\tau,3}$ & $(0.5,1.79)$ & $1.85$ & $15.64^\circ$ & $(0.5,1.41)$ & $1.50$ & $19.49^\circ$ \\
    \hline\hline
\end{tabular}
\end{table}
\begin{figure}[t]
	\centering
	\includegraphics[width = 0.98\linewidth]{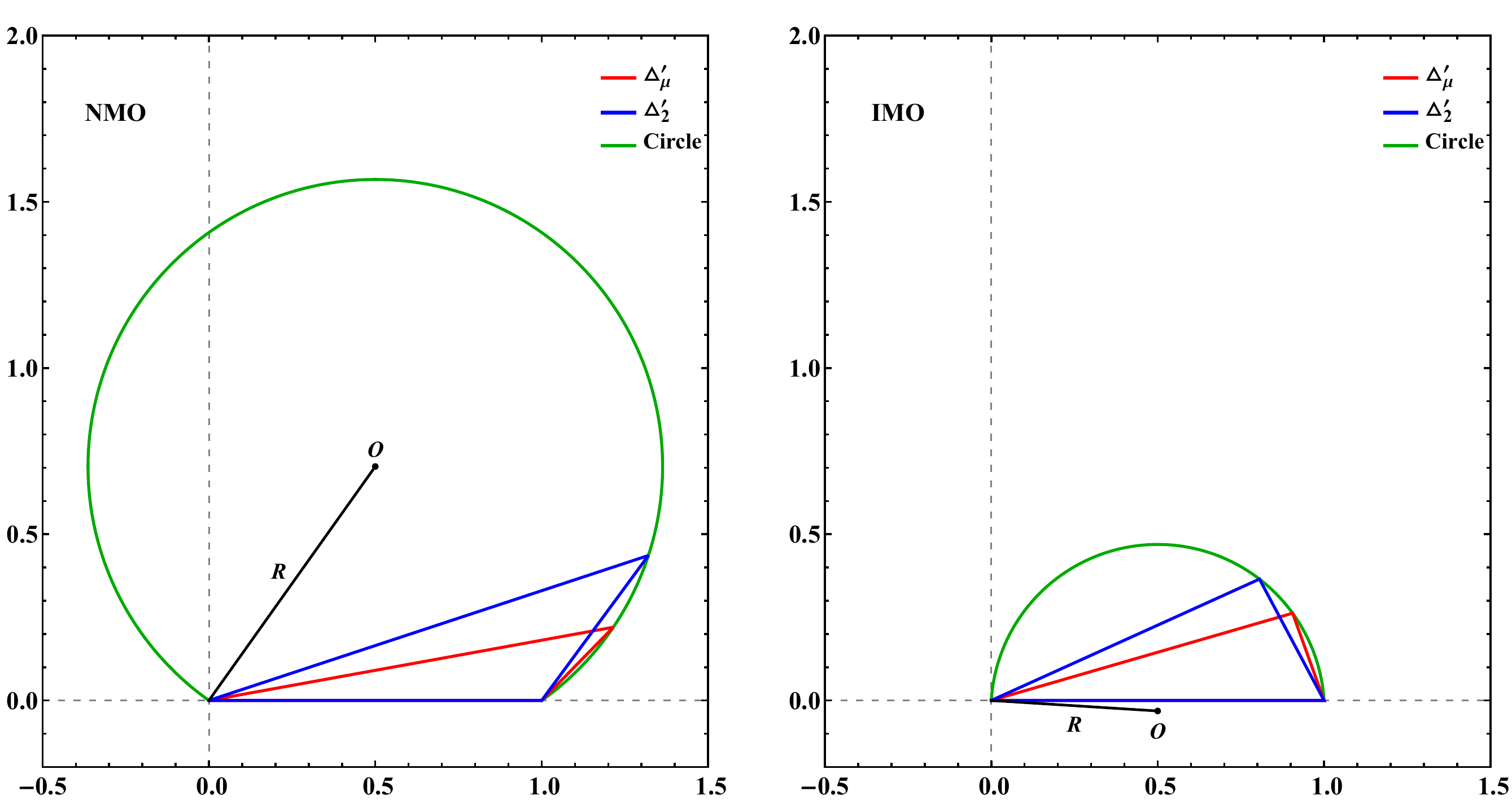}
    \vspace{0cm}
	\caption{The rescaled PMNS unitarity triangles $\triangle^\prime_\mu$ and $\triangle^\prime_2$
    in the complex plane for either the normal neutrino mass ordering (left panel) or the inverted
    mass ordering (right panel).}
\end{figure}
\begin{figure}[h!]
	\centering
	\includegraphics[width = 0.98\linewidth]{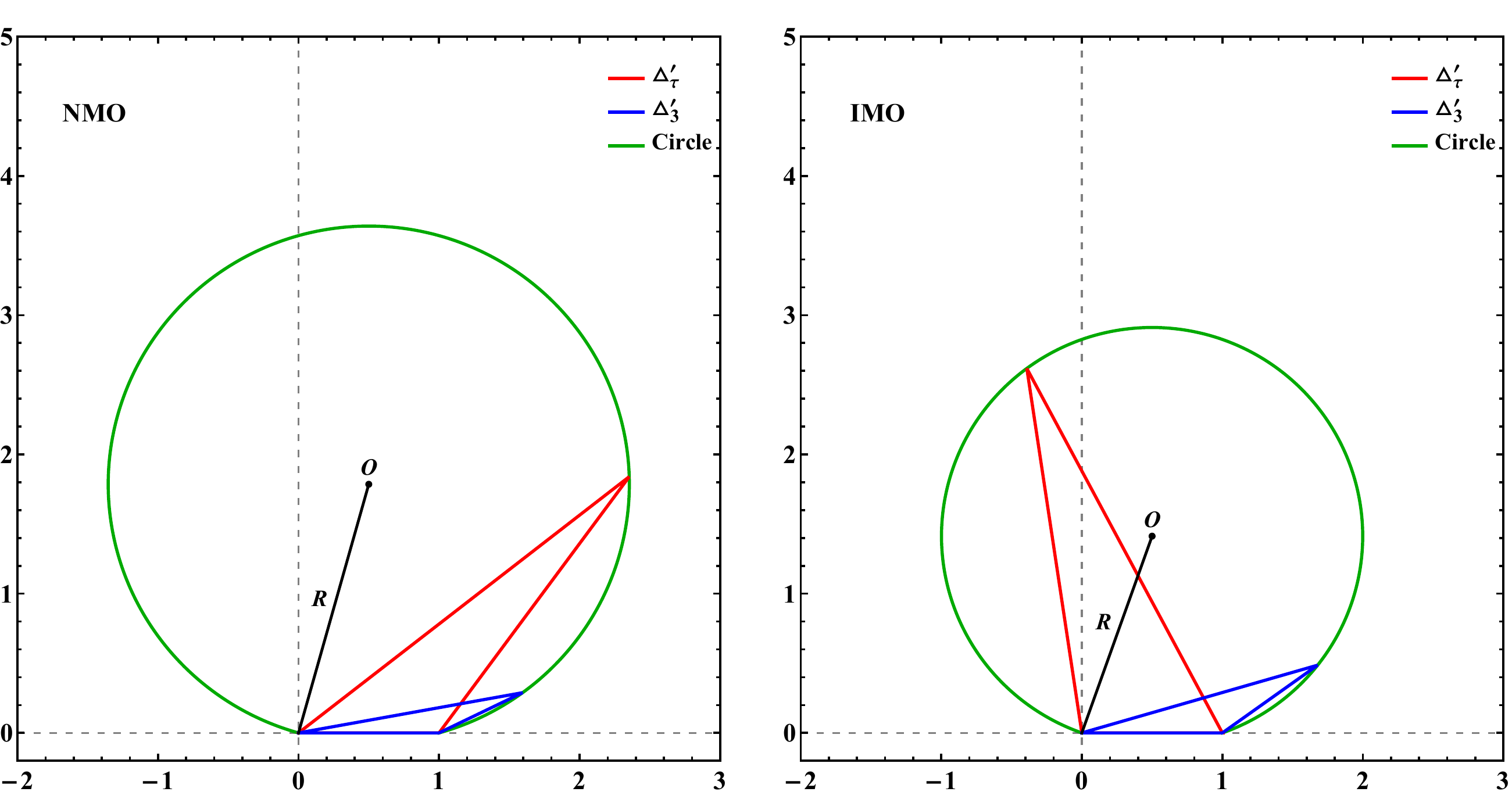}
    \vspace{0cm}
	\caption{The rescaled PMNS unitarity triangles $\triangle^\prime_\tau$ and $\triangle^\prime_3$
    in the complex plane for either the normal neutrino mass ordering (left panel) or the inverted
    mass ordering (right panel).}
\end{figure}

\section{Summary}

Motivated by the prospects that the {\it twin} CKM unitarity triangles $\triangle^{}_s$ and
$\triangle^{}_c$ will be precisely measured at the super-$B$ factory and the high-luminosity
LHC in the coming years, we address the question of whether they can be experimentally
distinguished from each other. With the help of the Wolfenstein parameters, we have calculated
the apex $(\overline{\rho}, \overline{\eta})$ of $\triangle^{\prime}_s$ and
the apex $(\widetilde{\rho}, \widetilde{\eta})$ of $\triangle^{\prime}_c$ --- the rescaled
versions of $\triangle^{}_s$ and $\triangle^{}_c$ in the complex plane --- and their
inner angles up to the accuracy of ${\cal O}(\lambda^6)$. In particular, we
find that the two apexes are actually located on a circular arc, whose
center and radius are given by $O = (0.5, 0.5 \cot\alpha)$ and $R = 0.5 \csc\alpha$
respectively. Both $\widetilde{\rho} - \overline{\rho}$ and
$\widetilde{\eta} - \overline{\eta}$ are found to be of ${\cal O}(\lambda^2)$, so
$\triangle^\prime_c$ and $\triangle^\prime_s$ are expected to be experimentally distinguishable if their apexes can be measured to a sufficiently good degree of
accuracy. We have pointed out that a similar experimental sensitivity will
allow us to probe small differences between the inner angles of $\triangle^\prime_c$ and
$\triangle^\prime_s$. We have also shown that $(\overline{\rho}, \overline{\eta})$
and $(\widetilde{\rho}, \widetilde{\eta})$ are insensitive to the two-loop RGE
running effects up to the accuracy of ${\cal O}(\lambda^4)$, and this observation
implies that the shapes of $\triangle^{}_c$ and $\triangle^{}_s$ keep invariant up to the
same accuracy when the energy scale $\Lambda$ changes between
$\Lambda^{}_{\rm EW} \sim 10^2~{\rm GeV}$ and $\Lambda^{}_{\rm GUT} \sim 10^{16}~{\rm GeV}$.
As a consequence, the experimental results of all the inner angles of
$\triangle^{}_c$ and $\triangle^{}_s$ obtained at low energies can directly be
confronted with some theoretical predictions at a superhigh energy scale.

As a by-product, three pairs of the rescaled PMNS unitarity triangles in the lepton
sector have been discussed in a similar but brief way. The apexes of each pair of
the triangles are also located on a circular arc, but we find that
their shapes look quite different if only the best-fit values of lepton flavor mixing
parameters are taken into account. Once the Dirac CP-violating phase $\delta$
is reliably determined in the upcoming long-baseline neutrino oscillation
experiments, it will be possible to systematically study the PMNS unitarity triangles
\footnote{Private communications with Shun Zhou.}.
In this respect we expect that the geometrical language under discussion may be
very helpful to test consistency of the PMNS picture for lepton flavor mixing
and CP violation and look for possible deviations caused by new physics associated
with the origin of tiny neutrino masses.

\section*{Acknowledgements}

One of us (Z.Z.X.) is deeply indebted to his schoolmate (A.J.) at Peking University for
motivating him to pay particular attention to all the twin phenomena in physics
and beyond since 1984. And we would like to thank Guo-yuan Huang for useful discussions.
This work was supported in part by the National Natural Science Foundation of China
under grant No. 11775231 and grant No. 11835013.


\end{document}